# Bilayer Stacking Ferrovalley without Breaking Time-Reversal Symmetry

Guoliang Yu[1,2], Junyi Ji[1,2], Changsong Xu[1,2]*, and H. J. Xiang[1,2,3]*

[1]*Key Laboratory of Computational Physical Sciences (Ministry of Education), Institute of Computational Physical Sciences, and Department of Physics, Fudan University, Shanghai 200433, China*

[2]*Shanghai Qi Zhi Institute, Shanghai 200030, China*

[3]*Collaborative Innovation Center of Advanced Microstructures, Nanjing 210093, China*

[†]*G.Y. and J.J. contributed equally to this work.*

Email: csxu@fudan.edu.cn, hxiang@fudan.edu.cn

**Abstract**

Non-volatile manipulation of valley polarization in solids has long been desired for valleytronics applications but remains challenging. Here, we propose a novel strategy for non-volatile manipulating valleys through bilayer stacking, which enables spontaneous valley polarization without breaking time-reversal symmetry. We call this noval physics as bilayer stacking ferrovalley (BSFV). The group theory analysis reveals that the two-dimensional (2D) valley materials with hexagonal and square lattices can host BSFV. By searching the 2D material database, we discovered 14 monolayer 2D materials with direct gaps that are candidates for realizing BSFV. Further first-principles calculations demonstrate that BSFV exists in $RhCl_3$ and InI bilayers. The bilayer stacking breaks their three- and four-fold rotation symmetry, resulting in 39 and 326 meV valley polarization, respectively. More interestingly, the valley polarization in our systems can be switched by interlayer sliding. Our study opens up a new direction for designing ferrovalley materials and thus greatly enriches the platform for the research of valleytronics.

**Main text**

Valleytronics is an emerging field that has gained much attention in recent years. It aims to utilize the valley degrees of freedom (DoF) of electrons in solids for information encoding and processing [1-6], instead of charge and spin DoF. Current research in valleytronics mainly focuses on two-dimensional (2D) systems, such as graphene and transition metal dichalcogenide (TMD) monolayers. [5-11]. Valleys in the crystals are degenerated due to some symmetry protections, for example, the K and K' valleys connected by time-reversal symmetry (TRs) in 2D hexagonal lattice [12], which hinders their application in valleytronics.

The prerequisite for using valley DoF as information carriers are to lift the valley degeneracy in these systems, leading to the so-called valley polarization. Numerous strategies have been proposed, such as optically pumping [13-15], applying a magnetic field [16-19] or electric field [20], but the resulting valley polarization is volatile. In contrast, non-volatile means are more compatible with the valleytronic device. One can support the valley materials with a magnetic substrate, which induces a magnetic proximity effect in the systems and thus breaks TRs [21-31]. Additionally, Duan *et al.* proposed the concept of 2D ferrovalley materials that exhibit spontaneous valley polarization due to TRs breaking caused by intrinsic magnetism [32-35]. Recently, several novel mechanisms have been proposed, such as ferroelectrically induced valley polarization [36-38]. The valleys of these systems are connected by a crystalline symmetry instead of TRs. Therefore, valley polarization can be induced by breaking the crystalline symmetry with spontaneous in-plane ferroelectricity. Despite these strategies achieve non-volatile valley polarization, they also have some disadvantages, such as in-plane magnetic easy axis and low critical temperatures of 2D ferromagnet [39], and the challenge to finding 2D in-plane ferroelectrics at room temperature [40]. Thus, there is still an urgent need to find an easy-to-implement and non-volatile means of achieve valley polarization.

Recently, vdW stacking has been proven to be an effective means of manipulating the properties of 2D materials. For example, magic-angle twisted bilayer graphene exhibits superconductivity [41,42] and 2D ferromagnets show stacking-dependent interlayer magnetic coupling [43-46]. By stacking two nonferroelectric monolayers, their inversion symmetry can be broken to realize ferroelectricity [47-52]. Moreover, our recent study propose a theory of bilayer stacking ferroelectricity and a general rules of the creation and annihilation of symmetries in bilayer stacking [52]. As valley degeneracy in a 2D system may be protected by crystalline symmetries, vdW stacking may serve as an efficient way to break crystalline symmetries so as to induce valley polarization.

In this Letter, we propose a concept of bilayer stacking ferrovalley (BSFV), a general strategy for manipulation of valleys in 2D systems. This strategy enables the easy realization of non-volatile valley polarization by bilayer stacking without breaking TRs, which goes beyond the existing paradigm of valleytronics. Through group theory analysis, we have found that this strategy can applied to 2D valley materials with hexagonal and square lattices, where the valleys are connected by non-time-reversal invariant (NTRI) points. By screening the material database, we discovered 14 kinds of 2D monolayer materials with direct band gaps suitable for this strategy. We further demonstrate the BSFV in $RhCl_3$ and InI bilayers by performing first-principles calculations. Breaking the $C_{3z}$ and $C_{4z}$ symmetry by interlayer stacking induces 39 and 326 meV valley polarization in $RhCl_3$ and InI bilayers, respectively. More interestingly, the valley polarization in these systems depends on the orientation of the interlayer sliding. Our results provide new possibilities for the research of ferrovalley.

***Design ideas and group theory analysis.*** The basic idea of realizing BSFV is to break the symmetries that map several degenerate valleys to each other in the Brillouin zone (BZ) of a two-dimensional (2D) system by bilayer stacking. For example, there is a symmetry operation that connects the two degenerate valleys to each other [see Fig.1 (c) and (e)]. We assume that the interlayer interaction is not too strong so that stacking without interlayer sliding will not change the positions of the valleys in the bilayer. If the connecting symmetry is broken by interlayer sliding, the valleys will be polarized [see Fig. 1 (d) and (f)].

Now we use the group theory to describe the valley polarization. We use $G$ and $G_{\boldsymbol{k}}$ to denote the layer group (LG) of a given 2D system and the group of the wave vector $\boldsymbol{k}$, respectively. For all $g_c$ in $G$ but not in $G_{\boldsymbol{k}}$, $E(\boldsymbol{k}) = E(g_c\boldsymbol{k})$ [53]. Therefore, the degenerated valleys locate at different wave vectors in the same $\boldsymbol{k}^*$ and are connected by one or several $g_c$. Bilayer stacking can theoretically break all point group symmetries but cannot change TRs. Thus, double degenerate valleys at $(k_x, k_y)$ and $(-k_x, -k_y)$, e.g., K and K′ in hexagonal crystal, cannot be polarized by bilayer stacking due to TRs.

Next, we discussed the conditions in four 2D crystal systems. 1) In the oblique crystal system, all valleys are connected by TRs [see SM]. 2) In the rectangular crystal system, only valleys not at the high symmetry lines or points can achieve valley polarization [see SM]. 3) In BZ of the square crystal system, the valleys located at $\Gamma(0,0)$ and $S\left(\frac{1}{2}, \frac{1}{2}\right)$ are single degenerate and cannot be polarized. Note the square crystal system must have four-fold rotation (or improper rotation), which will transform wave vectors except $\Gamma(0,0)$ and $S\left(\frac{1}{2}, \frac{1}{2}\right)$ into different wave vectors not connected by TRs. Therefore, the corresponding valleys can be polarized by breaking the four-fold rotation (or improper rotation). 4) In BZ of the hexagonal crystal system, the degeneracy of $\Gamma$ is 1 and K, K′ are connected by TRs. Hence, valleys located at the three points cannot be polarized. Note a hexagonal crystal system must have three-fold rotation, which will transform wave vectors except $\Gamma$, K, K′ into different wave vectors not connected by TRs. Thus, the corresponding valleys can be polarized by breaking the three-fold rotation. According to the group theory analysis of ref. [52], three-fold rotation $c_{3z}$, four-fold rotation $c_{4z}$, and four-fold improper rotation $s_{4z}$ can be broken by interlayer sliding, and the valley polarization can be realized (see Fig. 1(d) and (f)) and manipulated. We call this novel physics the bilayer stacking ferrovalley (BSFV).

To sum up, the valley possible to be polarized in the non-magnetic 2D crystals should be located and can only be located at a) trivial points in oblique systems, b) points except $\Gamma$ and S in square systems, and c) points except $\Gamma$, K, K′ in hexagonal systems. We focus on the high symmetry lines and points in square and hexagonal systems as shown in Fig. 1. It should be emphasized that valley polarization not only can be achieved by breaking $C_{3z}$, $C_{4z}$, and $S_{4z}$ in the square/hexagonal systems but also may be realized by breaking other symmetries (e.g., two-fold rotation $C_{2x}$). Moreover, whether the inversion symmetry is broken or not, the valley polarization can be realized. We listed all bilayer stacking and broken symmetries in square/hexagonal systems [see SM].

Based on the above analysis, we searched 652 non-magnetic 2D materials with a band gap larger than 0.2 eV in square or hexagonal crystal systems [54,55]. Then we selected desired ones with one of CBM or VBM located at high symmetry points and lines except $\Gamma, S, K, K'$. Finally, 352 valley materials are found to be potential BSFV materials. Among them, 14 and 338 valley materials are semiconductors with direct and indirect bandgap. All materials are listed in Table S1 in the Supplementary Material. First-principles calculations are performed on the hexagonal system $RhCl_3$ and the square system InI with direct bandgaps to demonstrate our predicted BSFV. It is worth noting that the BSFV concept can also be extended to these indirect bandgap valley materials. As an illustration, we have successfully applied the BSFV scheme to SnS bilayer system (as described in the Supplementary Material).

**BSFV in $RhCl_3$ with hexagonal lattice.** We applied the BSFV concepts to $RhCl_3$, a van der Waals (vdWs) material with a hexagonal lattice. Bulk $RhCl_3$ has a monoclinic C2/m structure and has been successfully prepared experimentally at room temperature [56]. Monolayer $RhCl_3$ has a crystal structure similar to $CrI_3$, which belongs to the P-31m space group and has $D_{3d}$ point group symmetry [57]. The unit cell consists of two Rh and six Cl atoms, in which the Rh atomic is sandwiched between two Cl atomic planes and forming a honeycomb lattice. We confirm that the monolayer is dynamically stable and has good thermal stability at room temperature by phonon spectra calculation and molecular dynamics simulation [see Fig. S1 in SM]. The electronic band structure shows that the monolayer $RhCl_3$ has a direct bandgap of about 1.72 eV with both the valence band maximum (VBM) and conduction band minimum (CBM) located at the M point [see Fig. S2 in SM]. A triple degeneracy of valleys occurs at the NTRI points $M_1$, $M_2$, and $M_3$, due to the three-fold rotation $C_{3z}$ symmetry connection [see Fig. S2]. Therefore, it is an excellent candidate for realize the concept of BSFV.

According to our concept, bilayer stacking can break the $C_{3z}$ symmetry in the hexagonal lattice and induce valley polarization. Next, we construct different $RhCl_3$ bilayers by stacking two monolayers. Here, using (*m,n*) to represent the interlayer sliding, that is, shifting the top layer by $ma_1+na_2$ relative to the bottom layer, where $a_1$ and $a_2$ represent lattice vectors. Further DFT calculations were performed to determine the energy minima for the different stacking configurations. As shown in Fig. 2(b), the top layer sliding by $D_1(1/3,2/3)$ and $D_2(2/3,1/3)$ results in a doubly degenerate global energy minimum, while shifting by $L_1(1/3,0)$, $L_2(0,1/3)$, and $L_3(2/3,2/3)$ leads to the triple degenerate local minima. The energy difference between local and global minima is 8 meV, which indicates good stability for local minima.

The space group and symmetry analysis of bilayer $RhCl_3$ with different stacking are shown that the G and $D_{1/2}$ stacking have $C_{3z}$ symmetry which is broken in the $L_{1/2/3}$ ones due to interlayer sliding. Therefore, it can be expected that the $L_{1/2/3}$ stacking can realize valley polarization, while the others do not. Below we focus on the electronic structure of bilayer $RhCl_3$ with G and $L_1$ stacking, which are shown in Figs. 2c and d. Those of all other configurations are shown in Fig. S3 since they show almost the same results as the G and $L_1$ stacking. The spin-orbit coupling (SOC) is not included because of its negligible strength. Here, we only focus on the band dispersion of the valley at the $M_1$, $M_2$, and $M_3$ points. For the bilayers with G stacking, the valleys at $M_1$, $M_2$, and $M_3$ points are energetically degenerate. When the top layer sliding $(1/3,0)$ leads to $L_1$ stacking, as expected, the energy degeneracy at $M_{1/2/3}$ points is lifted, resulting in a spontaneous valley polarization at the $M_2$ point. We define valley polarization as $\Delta^v_{valley} = E^v_{M_2} - E^v_{M_{1/3}}$ ($\Delta^c_{valley} = E^c_{M_2} - E^c_{M_{1/3}}$). Here, $v$ and $c$ denote the valence band and conduction band, $M_i$ represents the valley index. $\Delta^v_{valley}$ and $\Delta^c_{valley}$ are respectively 39 meV and 22 meV, which are comparable with those of $VSe_2$ (78 meV) [32], $LaBr_2$ (33 meV) [33], and $VSi_2N_4$ (63 meV) [35]. It is worth noting that the valleys at $M_1$ and $M_3$ points are degenerate due to the $m_{100}$ mirror symmetry connection. If the top layer sliding can cause the $m_{100}$ to break, for example $(1/6,1/3)$, the energy degeneracy at $M_1$ and $M_3$ points is removed [see Fig. S4]. Another important discovery is the valley polarization in our systems can be switched by interlayer sliding. Specifically, when the top layer slides $L_2(0,1/3)$ and $L_3(2/3,2/3)$ along the (010) and (110) directions, which results in a valley polarization at $M_1$ and $M_3$ points, respectively [see Fig. S3]. The kinetic pathways between different valley polarization states reveal that switching between $L_1$ and $L_2$ staking only needs to overcome a small energy barrier of 13 meV/f.u., and which requires $D_2$ as a transition state. The transformings between the $L_2$ and $L_3$ staking are similar, except that $D_1$ is the transition state [see Fig. S5]. Such results indicate that the switching between the valley polarization states $L_1$, $L_2$, and $L_3$ can be easily achieved. Therefore, $RhCl_3$ with hexagonal lattice can realize BSFV, and its valley polarization can be manipulated by interlayer translations.

***BSFV in InI with square lattice.*** Next, we applied our concepts in a square lattice using InI as an example. Bulk InI is a layered semiconductor material and has been synthesized in the experiments [58,59]. Recent research has demonstrated that monolayer and few-layer InI are as stable as their bulk counterparts [60,61]. Our band structure calculations show that the monolayer InI has two valleys of energy degeneracy on the high-symmetry path of $\Gamma - X$ and $\Gamma - Y$, which are connected by the four-fold rotation [see Fig. S6]. According to our group theory analysis, the four-fold rotation can be broken by interlayer sliding and lead to valley polarization. By performing DFT calculations of InI bilayers with different interlayer sliding, we find that the simply stacked bilayer exhibits no valley polarization and corresponds to the energy local minimum (see Fig. 3b and c). While the $N_1(1/2,0)$ and $N_2(0,1/2)$ stacking are the ground state of the system and achieve valley polarization of 326 meV at the CBM (see Fig. 3b and d). Such a large values are comparable to the ferroelectrically induced valley polarization of monolayer group-IV monochalcogenides [36]. This result in InI bilayer is consistent with our concept of BSFV.

Figure 2(b) further shows that there are two distinct switching paths between $N_1$ and $N_2$. The first path involves passing through G as a transition state and requires overcoming a barrier of 34 meV per formula unit. In contrast, the second path directly connects $N_1$ and $N_2$ and only requires overcoming smaller barriers of 25 meV per formula unit. Consequently, the second path is more favorable as it involves less energy expenditure, making it the more likely route for the system to follow during the switching process [see Fig. S7].

***Summary***. In summary, by group theory analysis of different 2D lattice systems, we identify that 2D hexagonal and square lattices may exist valleys connected by NRIP point. Based on this, we propose the concept of BSFV, which by bilayer stacking to achieve spontaneous valley polarization. Through searching the 2D material database, we discovered 14 candidates with direct band gaps for the realization of BSFV. Further performing first-principles calculations on $RhCl_3$ and InI bilayers, we demonstrate that the valley degrees of freedom in the hexagonal and square crystal lattice can be manipulated by different interlayer translations. Our discovery significantly expands the scope of valley polarization materials, which is crucial for the study of valleytronics.

**Acknowledgement**. This work is supported by NSFC (grants No. 811825403, 11991061, 12188101, 12174060, and 12274082) and the Guangdong Major Project of the Basic and Applied Basic Research (Future functional materials under extreme conditions--2021B0301030005). C.X. also acknowledge supports from the open project of Guangdong provincial key laboratory of magnetoelectric physics and devices (No. 2020B1212060030).

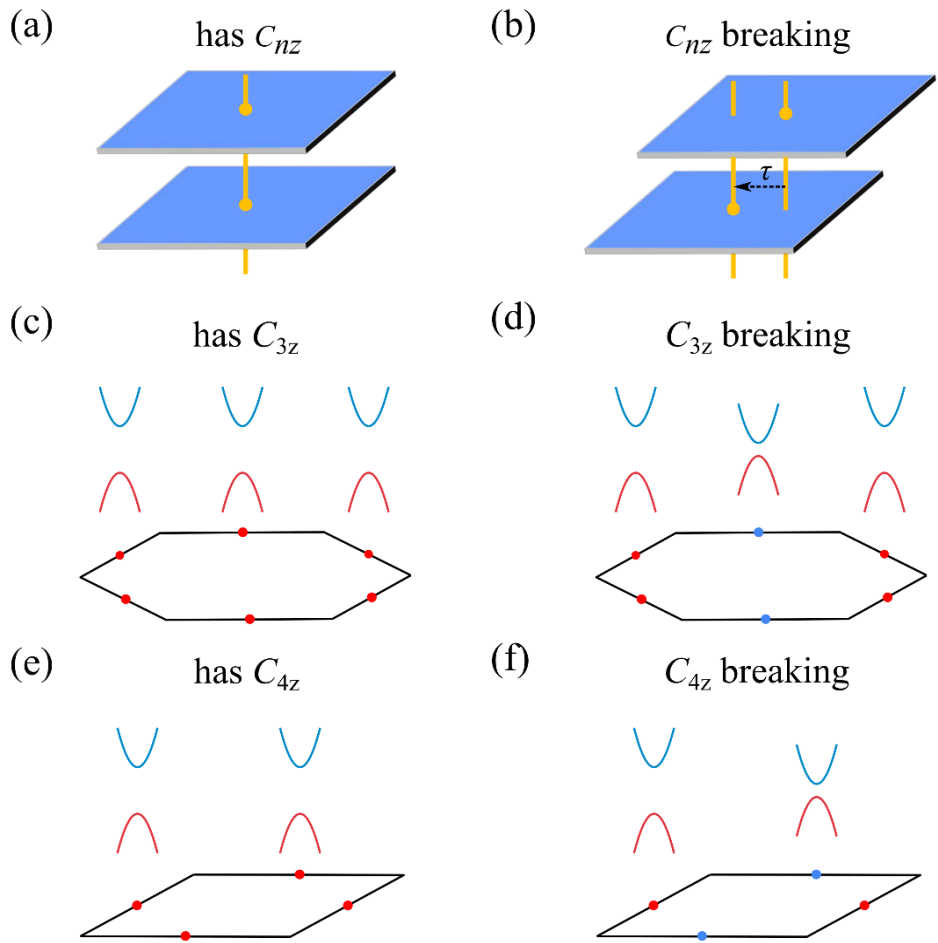

Figure 1. Stacking-induced valley polarization in vdWs bilayers. (a, b) Schematic diagrams of bilayers formed after direct stacking and relative sliding $\tau$ between two monolayers, respectively. Interlayer sliding may break the $C_{nz}$ symmetry of the system. Brillouin zone and valleys of a 2D hexagonal/square lattices (c)/(e) with and (d)/(f) without $C_{3z}/C_{4z}$ symmetry. Red and blue dots represent the degeneracy and polarized valleys, respectively. 2D hexagonal and square lattice has three and two degenerate valleys connected by $C_{3z}$ and $C_{4z}$ symmetry, respectively. Interlayer sliding breaks these symmetries leading to valley polarization.

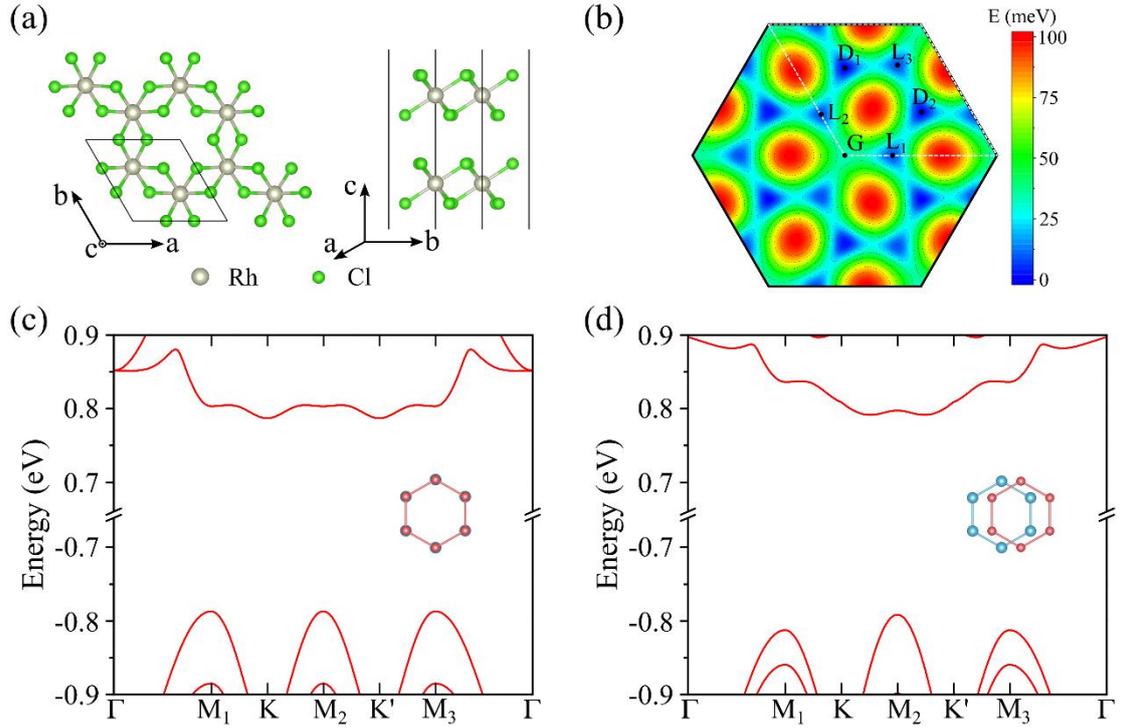

Figure 2. Bilayer stacking ferrovalley in RhCl3. (a) Geometric structures of bilayer RhCl3. The left and right panels show the top and side views of the directly stacked RhCl3 bilayers. Silver and green spheres denote Rh and Cl atoms, respectively. (b) Energy distribution for different bilayer stacking configurations. The white dotted parallelogram denotes the "unit cell". G, L1, L2, L3, M1, and M2 are six highly symmetrical translation points in the "unit cell", and their translation sizes are (0,0), (1/3,0), (0,1/3), (2/3,2/3), (1/3,2/3,), and (2/3,1/3). The global ground states around the original point are at M1 and M2 points, while L1/2/3 are the local minima. (c, d) Band structures for G and L1 points in (b), respectively, i.e., (0,0), (1/3,0). The inset shows the honeycomb lattice formed by Rh atoms in RhCl3 bilayers with different stacking, where the red and blue spheres denote the Rh atoms in the top and bottom layers, respectively.

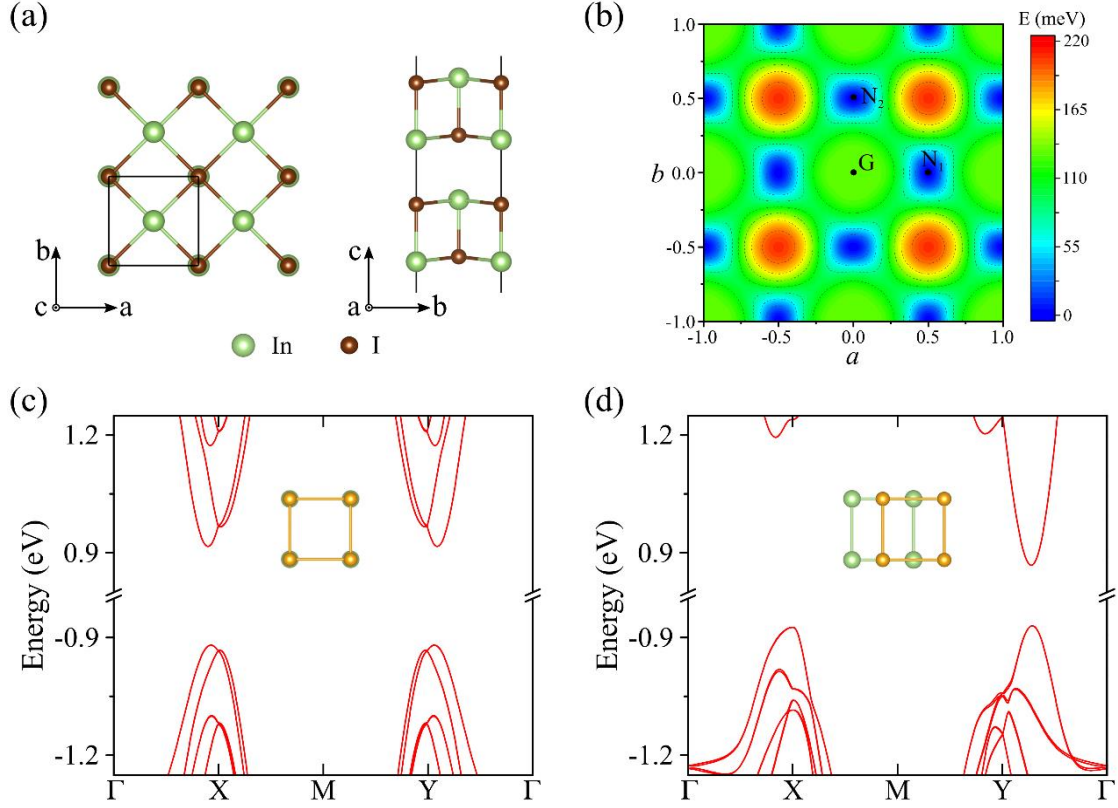

Figure 3. Bilayer stacking ferrovalley in a square lattice. (a) Top and side views of bilayer InI. Green and brown spheres denote In and I atoms, respectively. (b) Energy distribution of different translations between layers in a InI bilayer. *a* and *b* represent different lattice orientations, respectively. G, $N_1$, and $N_2$ are highly symmetrical stacking configurations with translation sizes (0,0), (1/2,0), and (0,1/2), respectively. $N_1$ and $N_2$ are doubly degenerate global minima, while G-stacking is a local minimum. (c, d) Band structures for G and $N_1$ points in (b), respectively, i.e., (0,0), (1/2,0). The insets show the positions of the In atoms in the upper layer in each layer in InI bilayers with G- and $N_1$-stacking, where the yellow and green spheres represent the In atoms in the top and bottom layers, respectively.